\documentclass[aps,prl,showpacs,twocolumn,superscriptaddress]{revtex}
\usepackage{graphicx}
\usepackage{dcolumn}
\usepackage{bm}
\usepackage[latin1]{inputenc}
\usepackage{amsfonts}
\usepackage{amssymb}
\usepackage{amsmath}
\usepackage{colordvi}
\usepackage{color}
\usepackage{ulem}

\begin{document}
\title{Position and Spin Control by Dynamical Ultrastrong Spin-Orbit Coupling}    

\author{C. Echeverr\'ia-Arrondo}
\affiliation{Department of Physical Chemistry, Universidad del Pa\'is Vasco UPV/EHU, 48080 Bilbao, Spain}
\author{E. Ya. Sherman}
\affiliation{Department of Physical Chemistry, Universidad del Pa\'is Vasco UPV/EHU, 48080 Bilbao, Spain}
\affiliation{IKERBASQUE, Basque Foundation for Science, Bilbao, Spain}
\date{\today}

\begin{abstract}
Focusing on the efficient probe and manipulation of single-particle spin states, 
we investigate the coupled spin and orbital dynamics of a spin 1/2 particle  
in a harmonic potential subject to ultrastrong spin-orbit interaction and external magnetic
field. The advantage of these systems is the clear visualization of the
strong spin-orbit coupling in the orbital dynamics. We also investigate 
the effect of a time-dependent coupling: Its nonadiabatic change causes 
an interesting interplay of spin and orbital motion which is related to the direction and magnitude
of the applied magnetic field. This result suggests that orbital state manipulation 
can be realized through ultrastrong spin-orbit interactions, becoming a useful tool for handling entangled 
spin and orbital degrees of freedom to produce, for example, spin desirable polarizations in time interesting for 
spintronics implementations.
\end{abstract}

\maketitle

Spin-orbit interactions have been proven very useful for realization of spintronics \cite{spintronics} 
with electrons in nanosystems.  
On one hand, it has been demonstrated in theory and experiment that spins can be 
tuned by various electric means.\cite{Rashba,Governale,Nowack,nature,Xuedong,Nowak} On the other hand, 
since spin-orbit coupling entangles spin and orbital motion, 
spin read-out is reachable by electric means.\cite{Levitov}
Such combination points to the possibility of probing 
and manipulating spins hosted by semiconductor quantum dots \cite{carlos,besombes} using only 
electric fields.\cite{nature} The spin-orbit control of qubits is a promising tool that suggests to
investigate the ultrastrong spin-orbit coupling regime to see all features of this technique.
Extreme spin-orbit interactions can be achieved at the surfaces of semiconductors coated with
heavy metals (see, e.g. [\onlinecite{Gierz,Yaji}]), which allow for spin manipulation 
by electric fields.\cite{Ibanez,Khomitsky12} Rashba performed a detailed analysis of two-dimensional 
quantum dots with the ultrastrong spin-orbit coupling [\onlinecite{Rashba12}]. 
Very recently, it was recognized that fully controllable strong interactions,
greatly beyond the range reachable in semiconductors,    
can be produced in ultracold atomic Bose and Fermi gases
by optical means.\cite{Wang,Cheuk} Similar to electrons in quantum dots, these systems are located in harmonic traps
and can demonstrate changeable in time spin-orbit
coupling, opening new venues for studies of related dynamics. 

\begin{figure}
\includegraphics*[scale=0.4]{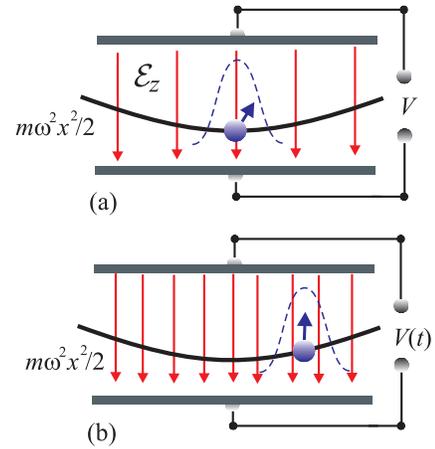}
\caption{(Color online) (a) Setup scheme for external generation of 
spin-orbit coupling through a bias voltage. Bold solid curves indicate the confining parabolic 
potential. (b) A non-adiabatic time-dependent field can cause electron displacement with spin rotation.}
\label{fig:fig1}
\end{figure}

The spin dynamics in these systems can be studied theoretically by 
analyzing the interactions between a harmonic oscillator and 
a two-level spin, making it similar to the  Jaynes-Cummings
model in quantum optics, as suggested by Debald and Emary.\cite{Debald}
This is a wide-purpose model (see e.g. \cite{Casanova,clarke,Braak}) applied in different fields of 
condensed matter and quantum optics such as
cavity quantum electrodynamics, trapped ions, 
and superconducting qubits (see [\onlinecite{Schmidt,Hu}] for recent results).
In addition, carbon nanotubes holding electron spins deeply  
coupled to the vibrational modes \cite{Palyi} can be described with the Jaynes-Cummings model.
 
The systems with spin-orbit couplings have several advantages 
not applicable elsewhere. We mention just two.
First, the coordinate dependence of the spin density 
makes it possible to visualize the effects of strong 
coupling in terms of particle position and measurable spin densities. 
Second, spin-orbit interaction 
and the Zeeman field can be made time-dependent \cite{field1,field2,field3} making 
the relevant dynamics both in spin and coordinate spaces accessible. 
These effects provide strongly nontrivial extensions of the conventional Jaynes-Cummings model. 

In a quantum dot, a spin-orbit strength $\alpha=\xi e\mathcal{E}_z$,
where $\xi$ is the material- and structure-dependent constant, can be 
induced by applying an external electric field $\mathcal{E}_z$,  \cite{Winkler}; 
a scheme for the experimental setup is given in Fig.~\ref{fig:fig1}. In cold atomic gases
this time-dependent modification can be reached by changing amplitudes and geometries of the 
corresponding laser fields. 

In this Letter, we present a description of the spin dynamics of an electron in a semiconductor quantum 
dot or confined cold atoms, both systems subject to strong spin-obit interactions.
We focus on a one-dimensional harmonic oscillator 
with spin degree of freedom, capturing main physics of the systems of interest. 
This oscillator is 
under an applied magnetic field which one can rotate with respect to the coordinate axes.  
We consider two types of spin-orbit couplings, constant and time-dependent; the results given below thereby 
acquiring wider application. 

The eigenstates can be obtained from the following Hamiltonian: 
\begin{equation}
\hat H(t)=\hbar^2k^2/2m+m\omega^2x^2/2+\alpha(t)\sigma_{x}k+\frac{\Delta}{2}\sigma_{\Delta}, 
\end{equation}
where $m$ is the particle effective mass,
$\Delta$ and $\theta$ are the Zeeman splitting and tilt angle of the magnetic field as applied in $xz$ plane, 
respectively, $\sigma_{\Delta}$=$\sigma_x\sin\theta+\sigma_z\cos\theta$ is the 
corresponding spin projection, and $\sigma_{x},\sigma_{z}$ are the Pauli matrices. 
In the second quantization this Hamiltonian reads
\begin{equation}
\hat H=\hbar\omega (\hat a^{\dag} \hat a + 1/2)+i\hbar\omega g(t)(\hat a^{\dag} -\hat a)\sigma_x + \frac{\Delta}{2}\sigma_{\Delta},
\end{equation}
where $\hat a^{\dag}$ and $\hat a$ are the creation and annihilation operators, respectively,
and $g$=$\alpha\sqrt{m/2\hbar^{3}\omega}$ is a dimensionless coupling constant, which
can be understood as the ratio of the characteristic anomalous spin-dependent velocity $\alpha/\hbar$ to 
the characteristic quantum velocity spread in the ground state of the harmonic oscillator $\sqrt{\hbar\omega/m}$,
or as the ratio of the quantum oscillator length $l_0=\sqrt{\hbar/m\omega}$ to the spin precession length $\hbar^{2}/m\alpha.$ 
We use the basis of spin orbitals $|n\rangle|\sigma\rangle$, composed of the eigenstates 
of $\hat{a}^{\dag}\hat{a}$, $|n\rangle$, and those of 
$\sigma_{z}$, $|\sigma\rangle\equiv\left|\uparrow\right>_{z}$ 
and $\left|\downarrow\right>_{z}$ with respect to the $z$-axis. Numerical values of $g$ strongly vary
from system to system. For InSb-based quantum dots, where $\alpha$ can reach 
$10^{-5}$ meVcm, $m$ is of the order of 0.02 of the free electron mass, and $\omega\sim10^{12}$ s$^{-1}$, 
one can expect $g\approx1$. For cold fermions such as $^{40}$K, where $\alpha/\hbar$ 
can be of the order of $10$ cm/s and $\omega\sim10^{3}$ s$^{-1}$, $g$ can be of the order of 10. 
 
To make connection to previous works on the ultrastrong regime (see, e.g. [\onlinecite{Casanova}]), 
first we investigate the effect of a constant coupling. 
The eigenstates of the full Hamiltonian $|\phi_i\rangle$ 
have the form $\sum_{n}|n\rangle(c_n^{\rm{u}}\left|\uparrow\rangle_{z}\right.+c_n^{\rm{d}}\left|\downarrow\rangle_{z}\right.)$
with the expansion coefficients $c_n^{\rm{u}}$ and $c_n^{\rm{d}}$; 
the normalized orbitals are expressed in the $x$-representation as
\begin{equation}
\langle x|n\rangle=\root 4\of {\frac{1}{\pi l_0^2 2^{2n}(n!)^2}}\exp\left[-\frac{x^2}{2l_0^2}\right]H_n\left[\frac{x}{l_0 }\right],
\end{equation}
where $H_n$ is the $n$-th order Hermite polynomial. In the limit of a very weak coupling, $g\ll 1$, in an arbitrarily
directed magnetic field,  $|\phi_i\rangle$ contains five main contributions. The main one is the 
direct product of $|n\rangle$ and eigenstate of $\sigma_{\Delta}$. The other four are perturbative terms
of direct products of $|n\pm1\rangle$ and eigenstates of $\sigma_{\Delta}$. At $\theta=0$, corresponding
to the conventional Jaynes-Cummings model, spin selection rules exclude two of these four states. 
Moreover, at $\theta=0$, in the eigenstates the contributions with different spatial parities have opposite spins. 
The expectation value of the velocity in the eigenstates is 
zero, $\langle v\rangle$=$d\langle x\rangle/dt$=0; however, the mean momentum is finite:
\begin{eqnarray}
&&\langle v\rangle= \frac{i}{\hbar}\langle\left[H,x\right]\rangle=
\frac{\hbar}{m}\langle k\rangle + \frac{\alpha}{\hbar}\langle\sigma_x\rangle=0, \nonumber \\
&&\langle k\rangle=-\sqrt{2}\frac{g}{l_{0}}\langle\sigma_x\rangle.
\label{vkmean}
\end{eqnarray}
Correspondingly, for the coordinate $\langle\phi_i|x|\phi_i\rangle=0$. Equation (\ref{vkmean}) corresponds to
zero mechanical momentum $\hbar k-{\cal A}$ for the gauge \cite{Aleiner} ${\cal A}=-m\alpha\sigma_{x}/\hbar$.
The total spectrum results from the magnetic-field mixing of two parabolic branches, 
that for $\langle k\rangle=-\sqrt{2}{g}/l_{0}$ when the spin state is $\left|\uparrow\rangle_{x}\right.$, 
and that for $\langle k\rangle=\sqrt{2}{g}/l_{0}$ when it is $\left|\downarrow\rangle_x\right.$.

\begin{figure}[t]
\includegraphics*[scale=.85]{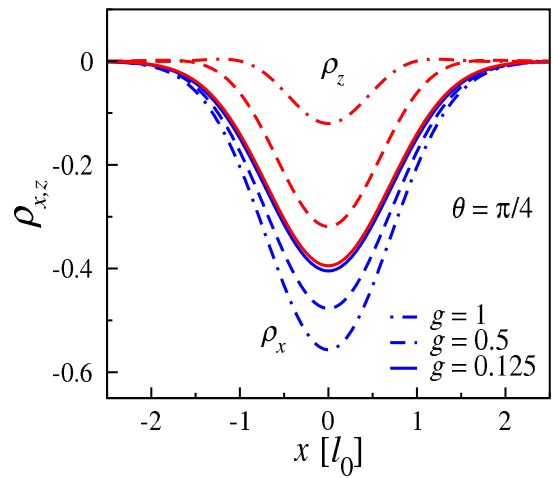}
\caption{(Color online) Nonzero spin densities $\rho_{x,z}$ for 
different $g$ values when the electron is at the ground state. 
Note that $\rho_x$ and $\rho_z$ merge for 
smaller $g$ coupling. Here and below we use a
truncated Hilbert space
of 64 states, sufficient to study ultrastrong spin-orbit coupling.
The Zeeman splitting in all calculations is taken as $\Delta=0.5\hbar\omega$. }
\label{fig:fig2}
\end{figure}

To show the advantages of systems with
spin-orbit coupling, we calculate the spatially resolved spin 
densities $\rho_j(x)$, $j=x,y,z$, providing valuable info about
the system.\cite{Khomitsky} For an arbitrary state $|{\bm \psi}\rangle$, 
presented in the form $\sum_{n}|n\rangle(a_n^{\rm{u}}\left|\uparrow\rangle_{z}\right.+a_n^{\rm{d}}\left|\downarrow\rangle_{z}\right.)$, these functions are defined as 
\begin{equation}
\rho_j(x)=\sum_{n,m}\langle n|x\rangle\langle x|m\rangle (a_n^{\rm{u}\ast},a_n^{\rm{d\ast}})\sigma_j 
\begin{pmatrix}
a_m^{\rm{u}} \\
a_m^{\rm{d}}
\end{pmatrix}.
\end{equation}
We focus on a particle in the ground state and take $\theta$=$\pi/4$ as an example. 
The densities $\rho_j(x)$  are presented in Fig.~\ref{fig:fig2}. The integrals of $\rho_{j}(x)$ 
over the $x$-coordinate are the spin expectation values. 

\begin{figure}
\includegraphics*[scale=0.62]{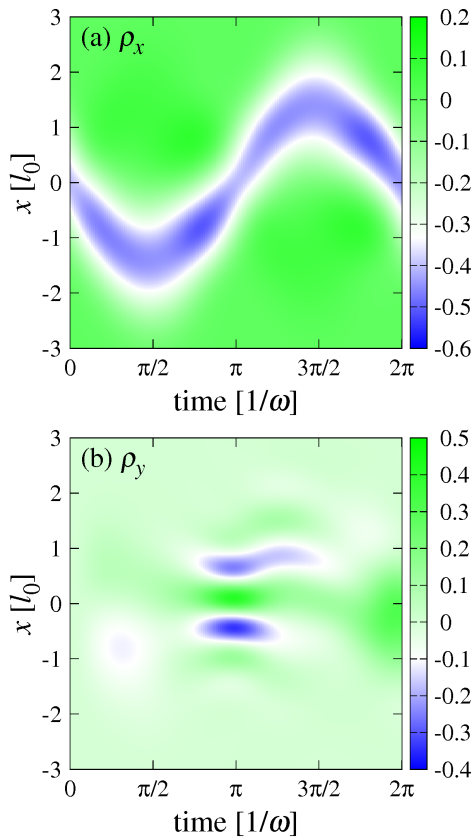}
\caption{(Color online) Spin densities (a) $\rho_x$ and (b) $\rho_y$ as a 
function of time for an electron spin antiparallel to the applied magnetic field with 
$\theta=\pi/4$; here we take the marginal case $g$=1.}
\label{fig:fig3}
\end{figure}

Next, we analyze the coupled dynamics of a system put away from an eigenstate. 
For this purpose, we choose as a typical example an eigenstate of $\sigma_{\Delta}$ antiparallel 
to the magnetic field: 
\begin{equation}
\label{psi0}
|{\bm \psi}(0)\rangle=|0\rangle
(-\sin(\theta/2)\left|\uparrow\rangle_{z}\right.+\cos(\theta/2)\left|\downarrow\rangle_{z}\right.),
\end{equation}
The time dependence is obtained from $|{\bm\psi}(t)\rangle=\sum_{i}\zeta_{i}|\phi_{i}\rangle e^{-iE_it/\hbar},$
where $\zeta_{i}$ are the corresponding expansion coefficients. 
We study the dynamics of spin densities for $\theta=\pi/4$ and $g=1$.
The particle oscillations are shown in Fig.~\ref{fig:fig3}(a). The 
Gaussian-like shape of $\rho_x$ is robust against time; however, those of $\rho_y$ 
and $\rho_z$ (not shown in the Figure) are fully changed. As a consequence, 
we see a strong correlation between spin state and position of the particle even
in the dynamical regime.  

One of the main advantages of quantum dots and cold atoms is the ability to
manipulate the strength of spin-orbit coupling and, thus, to cause dynamics in the orbital
and spin channels. The time-dependent spin-orbit coupling can be used for generation of 
spin currents in two-dimensional electron gas \cite{Malshukov} and spin separation in two-electron quantum dots 
similar to that predicted in Ref.[\onlinecite{Szafran13}].  
Here we take a single-period perturbation $g(t)=g_{0}\sin(\Omega t)$ at frequency $\Omega$,
fast enough to yield an appreciable nonadiabatic behavior, but
being sufficiently slow to allow using the 
available electrical and optical means to generate the spin-orbit interaction. 
To clearly see the effects of strong spin-orbit coupling, we compare below the obtained
numerically exact result with the perturbation theory.

To use the perturbation theory we take the basis of first four eigenstates 
\begin{equation}
\psi_{1}=|0\rangle|\downarrow\rangle_{z},\hspace{0.2cm} \psi_{2}=|0\rangle|\uparrow\rangle_{z}, 
\hspace{0.2cm}\psi_{3}=|1\rangle|\downarrow\rangle_{z},
\hspace{0.2cm}\psi_{4}=|1\rangle|\uparrow\rangle_{z}
\label{basis}
\end{equation}
where time-dependent wavefunction becomes
\begin{equation}
{\bm \psi} (t)=a_{1}(t)\psi_{1} +a_{3}(t)\psi_{3}e^{-i\omega t}
+a_{4}(t)\psi_{4}e^{-i\left( \omega +\Delta \right) t}.
\label{psi:t} 
\end{equation}
The assumed time-dependence yields the expansion coefficients 
\begin{eqnarray}
&&a_{3}(t)=-g_{0}\sin \mathrm{\theta }\int_{0}^{t}\sin (\Omega \tau
e^{i\omega \tau }d\tau \\
&&a_{4}(t)=g_{0}\cos \mathrm{\theta }\int_{0}^{t}\sin (\Omega \tau )e^{i\left( \omega +\Delta \right) \tau }d\tau. 
\label{a:t} 
\end{eqnarray}

The expectation values of coordinate and spin projection onto the magnetic are expressed as:
\begin{eqnarray}
\langle x(t)\rangle\equiv
\left\langle {\bm \psi} (t)\right| \widehat{x}\left| {\bm \psi} (t)\right\rangle =
\frac{1}{\sqrt{2}}\left( a_{3}(t)e^{-i\omega t}+\mbox{c.c.}\right) \label{timedependencex} \\
\langle \sigma_{\Delta}(t)\rangle\equiv
\left\langle {\bm \psi} (t)\right|\sigma_{\Delta}\left| {\bm \psi}
(t)\right\rangle =-1+2\left|a_{4}^{2}(t)\right|.
\label{timedependence}
\end{eqnarray}
These perturbation theory formulas show the role of the direction  of magnetic
field on the spin and spatial dynamics, not present in the conventional Jaynes-Cummings model
and allowing to extend the abilities for coordinate and spin manipulation. 

\begin{figure}[t]
\includegraphics*[scale=0.85]{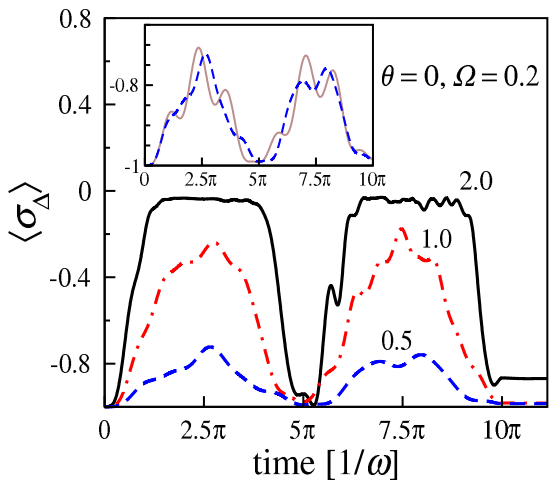}
\caption{(Color online)  Dynamics of spins under a
time-dependent $g(t)$=$g_0\sin(\Omega t)$. Lines are marked by 
corresponding $g_0$ values. Inset shows comparison of the exact (dashed line) and perturbation theory (solid line) results for
$g_{0}=0.5$.}
\label{fig:fig4}
\end{figure}

We treat the problem numerically for a single period $T=2\pi/\Omega$ [\onlinecite{Runge}],  
beginning with $\theta=0$, where Zeeman and spin-orbit fields are orthogonal, similar to the
Jaynes-Cummings model. Figure \ref{fig:fig4} demonstrates time dependence of $\langle\sigma_{\Delta}(t)\rangle$ 
for different couplings $g_{0}$ and comparison with the perturbation result for $g_{0}=0.5$ in the inset.
As it can be seen in Fig.~\ref{fig:fig4}, spin projection at the magnetic field changes strongly with
time, corresponding to the spin rotation due to the spin-orbit coupling,
and remains constant, as expected, after the change stops. The value of this projection after the end of
the perturbation corresponds to the degree of nonadiabaticity of this process. 
It is interesting to mention the appearance of plateaus at strong spin-orbit coupling regime.
These plateaus show that even a low-frequency dynamics is strongly nonadiabatic. The reason is the 
following. In the presence of spin-orbit coupling and magnetic field with $\theta=0$, the splitting of the 
ground-state doublet, $\Delta\exp(-2g^{2})\ll\Delta$, is small due to weakly overlapping eigenstates
in the momentum space, as discussed after Eq.(\ref{vkmean}).  As a result, even very slow changes in the system 
parameters cannot be treated adiabatically.  Here spatial motion  does not occur ($\langle x(t)\rangle=0$), 
in agreement with Eq.(\ref{timedependencex}).

\begin{figure}[t]
\includegraphics*[scale=0.85]{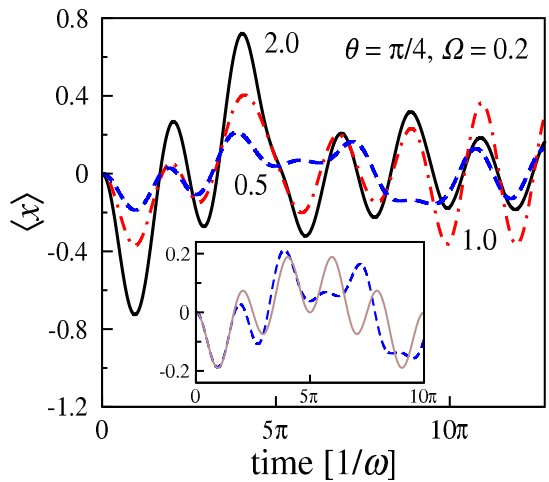}
\caption{(Color online)  Dynamics of particle mean coordinate under 
time-dependent strength $g(t)$=$g_0\sin(\Omega t)$. Lines are marked by 
corresponding $g_0$ values. 
Inset shows comparison of the exact (dashed line) and perturbation theory (solid line) results for
$g_{0}=0.5$.}
\label{fig:fig5}
\end{figure}

\begin{figure}[t]
\includegraphics*[scale=0.85]{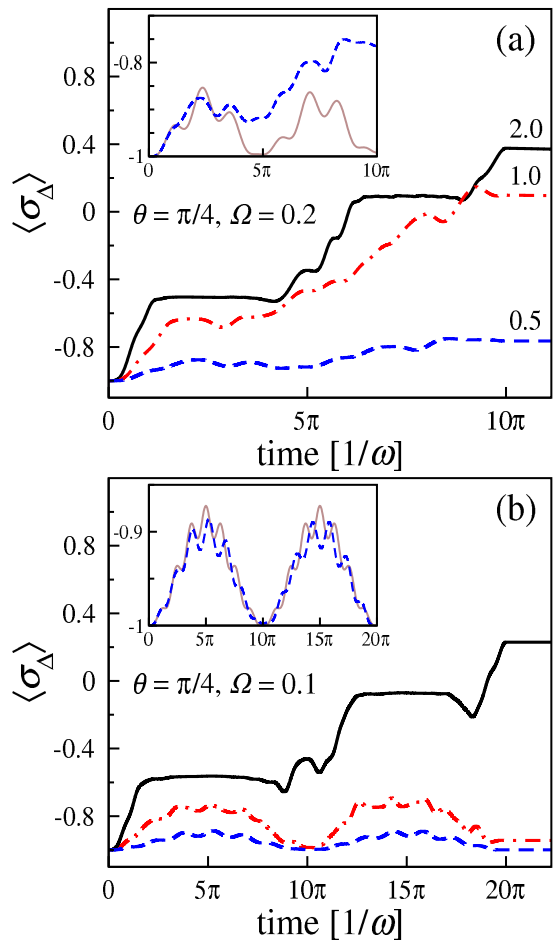}
\caption{(Color online) Dynamics of $\langle\sigma_{\Delta}\rangle$ under 
time-dependent strength $g(t)$=$g_0\sin(\Omega t)$. Lines are marked by 
corresponding $g_0$ values. The initial state is the same as in Fig.\ref{fig:fig5}.
Insets show comparison of the exact (dashed line) and perturbation theory (solid line) results for
$g_{0}=0.5$. (a) $\Omega=0.2$, (b) $\Omega=0.1$.}
\label{fig:fig6}
\end{figure}

Next, we take the state of Eq.(\ref{psi0}), with $\theta=\pi/4$ as the initial one
to demonstrate the qualitative role of the magnetic field direction.
First qualitative effect is nonzero oscillations of the coordinate, 
nonmonochromatic at $t<T$, as depicted in Fig.~\ref{fig:fig5}. 
The oscillations in $\langle x(t)\rangle$ are caused mainly by transitions between 
ground and first excited eigenstates of the model Hamiltonian; these states get mixed by 
the strong coupling. The amplitude of oscillations increases with $g_0$, as expected, 
but also with $\Omega$. The behavior strongly depends on the frequency and amplitude of spin-orbit
coupling, as can be seen by comparison of adiabatic and nonadiabatic plots. 
After the end of perturbation, the coordinate oscillates 
at the frequency $\omega$. Next qualitative difference is in the 
behavior of $\langle \sigma_{\Delta}(t)\rangle.$   
Here, in the case of a strong static spin-orbit coupling, the splitting of the
lowest doublet is $\Delta\cos\theta$. 
We see the effects of nonadiabatic perturbation and strong coupling in Fig.~\ref{fig:fig6}.
Dependent on $g_{0}$ and $\Omega$, different regimes in the dynamics of $\langle \sigma_{\Delta}(t)\rangle$
are possible: the system either returns to the initial state after the end of perturbation or 
switches to other states. The switching effect depends on the frequency of the field. 
At a small frequency, the dynamics is more adiabatic and perturbative, while for 
the larger frequency the switching can occur. 
The inset shows that the perturbation theory works at short time, while 
at longer times more states participate, and the results become different. From this
plot we can formulate the condition of strong spin-orbit coupling for a given system as
a transition from peak-like (``weak'' or ``moderate'' coupling) to step-like (``strong'') 
evolution of $\langle \sigma_{\Delta}(t)\rangle$. It is interesting to mention that
the transition and, therefore, the definition of the coupling strength in a dynamical system 
depends on the frequency of the applied external field.

In summary, we have investigated how strong dynamical spin-orbit couplings 
can be applied to probe and manipulate spins of electrons in semiconductor 
quantum dots and cold atoms in parabolic confinement through the correlated
spin and orbital motion. We reveal the importance of the tilt angle of the applied 
magnetic field, the effect strongly beyond the conventional Jaynes-Cummings
model. The obtained dynamics shows that, under a strong constant coupling, 
a particle oscillates in correlation with its spin orientation. The motion 
of the particle can be influenced by time-dependent coupling with the result
strongly dependent on all parameters. We observe a transition from periodic
to step-like behavior of spin component parallel to the magnetic
field with increasing the coupling strength.
This fact clarifies the way to define qualitative effect if the ultrastrong spin-orbit
coupling. The present work widens the applicability of the spin-orbit control, 
as it covers different strengths for the induced interaction and tilt angles for 
the applied magnetic field. It also emphasizes the usability of electric and optical
fields for spin probe and manipulation, which are crucial for spintronics. 
These results may also be of interest for quantum optics and quantum information realizations.

We gratefully acknowledge fruitful discussions with  E. Il'ichev, G. Romero, E. Solano,
and, especially, with J. C. Retamal. We acknowledge support of the MINECO of Spain 
(grant FIS 2009-12773-C02-01), the Government of the Basque Country (grant
"Grupos Consolidados UPV/EHU del Gobierno Vasco" IT-472-10), and the UPV/EHU (program UFI 11/55).



\begin{thebibliography}{}
\bibitem{Runge} The dynamics are computed using a 4$^{\rm{th}}$-order 
Runge-Kutta method till $t=2\pi/\Omega$, and using the  
decomposition $|{\bm \psi}(t)\rangle=\sum_{i}\zeta_{i}|\phi_i\rangle e^{-iE_it/\hbar}$ 
beyond this time.
\end{thebibliography}

\end{document}